\documentclass[prd,twocolumn,showpacs,superscriptaddress,preprintnumbers,amsmath,amssymb]{revtex4-1}

\usepackage{graphicx} % Required for inserting images
\usepackage{tikz}
\usetikzlibrary{arrows}

\usepackage[colorlinks=true,urlcolor=blue,linkcolor=blue,citecolor=blue]{hyperref}
\usepackage[normalem]{ulem}
\usepackage[capitalize]{cleveref}
 
\usepackage{comment}
\usepackage{flushend}

\usepackage[T1]{fontenc} % if needed

\usepackage{mathtools}

\usepackage{xcolor}
\usepackage{graphicx}% Include figure files
\usepackage{dcolumn}% Align table columns on decimal point
\usepackage{bm}% bold math
\usepackage{multirow}

\usepackage{epstopdf}

\newcommand{\Beq}{\begin{equation}\begin{aligned}}
\newcommand{\Eeq}{\end{aligned}\end{equation}}

\begin{document}

\preprint{IPMU23-0008, KEK-QUP-2023-0006, KEK-TH-2514, KEK-Cosmo-0308, YITP-23-44}

\title{Universal Gravitational Wave Signatures of Cosmological Solitons} 

\author{Kaloian D. Lozanov}
\email{kaloian.lozanov@ipmu.jp}
\affiliation{Kavli Institute for the Physics and Mathematics of the Universe (WPI), UTIAS
The University of Tokyo, Kashiwa, Chiba 277-8583, Japan.}

\author{Misao Sasaki}
\email{misao.sasaki@ipmu.jp}
\affiliation{Kavli Institute for the Physics and Mathematics of the Universe (WPI), UTIAS
The University of Tokyo, Kashiwa, Chiba 277-8583, Japan.}
\affiliation{Center for Gravitational Physics and Quantum Information, Yukawa Institute for Theoretical Physics,
Kyoto University, Kyoto 606-8502, Japan}
\affiliation{Leung Center for Cosmology and Particle Astrophysics, National Taiwan
University, Taipei 10617, Taiwan}

\author{Volodymyr Takhistov}
\email{vtakhist@post.kek.jp}
\affiliation{International Center for Quantum-field Measurement Systems for Studies of the Universe and Particles (QUP, WPI),
High Energy Accelerator Research Organization (KEK), Oho 1-1, Tsukuba, Ibaraki 305-0801, Japan}
\affiliation{Theory Center, Institute of Particle and Nuclear Studies (IPNS), High Energy Accelerator Research Organization (KEK), Tsukuba 305-0801, Japan
}
\affiliation{Graduate University for Advanced Studies (SOKENDAI), \\
1-1 Oho, Tsukuba, Ibaraki 305-0801, Japan}
\affiliation{Kavli Institute for the Physics and Mathematics of the Universe (WPI), UTIAS
The University of Tokyo, Kashiwa, Chiba 277-8583, Japan.}

\date{\today}

\begin{abstract}
Cosmological solitonic objects such as monopoles, cosmic strings, domain walls, oscillons and Q-balls often appear in theories of the early Universe. We demonstrate that such scenarios are generically accompanied by a novel production source of gravitational waves stemming from soliton isocurvature perturbations.
The resulting induced \textit{universal gravitational waves} (UGWs) reside at lower frequencies compared to gravitational waves typically associated with soliton formation. 
We show that UGWs from axion-like particle (ALP) oscillons, originating from ALP misalignment, extend the frequency range of produced gravitational waves by more than two orders of magnitude regardless of the ALP mass and decay constant and can be observable in upcoming gravitational wave experiments.
UGWs open a new route for gravitational wave signatures in broad classes of cosmological theories.
\end{abstract}

\maketitle

{\it Introduction}.--- 
Aside from particles, long-lived non-linear field configurations known as solitons often appear in physical theories, ranging from those describing condensed matter systems to various extensions of the Standard Model of elementary particles~(e.g.~\cite{Weinberg:2012pjx}). 
The extreme high temperature conditions of the early Universe are ideal for the formation of overdense cosmological solitonic objects prior to the epoch of Big Bang nucleosynthesis \cite{Kibble:1976sj,Hindmarsh:1994re}. 

Depending on the symmetry properties of matter and the details of a potential cosmological phase transition, distinct types of cosmological solitons, including monopoles (dimension 0), cosmic strings (dimension 1), domain walls (dimension 2), associated with topologically non-trivial vacuum, can appear via the Kibble-Zurek mechanism~\cite{Vilenkin:2000jqa}. Depending on details of scenarios, such topological defects can play the role of dark matter (DM)~(e.g.~\cite{Murayama:2009nj}).
Theories with attractive self-interactions of quantum fields can also support long-lived spatially localized solitonic overdensities. 
Real scalar fields fragmenting into oscillon ``lumps'' commonly occur in models of inflation favored by observations~\cite{Amin:2011hj,Lozanov:2017hjm}.
Fragmentation of a complex scalar field into Q-balls are a typical prediction of theories aiming to explain the observed baryon-anti-baryon asymmetry of the universe~(see e.g.~\cite{Allahverdi:2012ju}). 
Variety of phenomena, such as formation of primordial black holes~\cite{Cotner:2016cvr,Cotner:2018vug,Cotner:2019ykd}, have been associated with such cosmological solitons.
While unreachable with energies accessible in laboratories and accelerators, cosmological solitons are intimately connected with the early Universe and their signatures offer unique probes of new physics.

Solitons form in the primordial Universe through rapid non-linear processes on scales smaller than the cosmological horizon. The brief evolution of the inhomogeneities is an efficient source for production of gravitational waves (GWs), as well exemplified by oscillon formation~\cite{Weir:2017wfa,Dufaux:2010cf,Lozanov:2019ylm}. The resulting single-peak stochastic GW background has a central frequency determined by the non-linear scale \cite{Weir:2017wfa,Amin:2014eta}. Such GW backgrounds are a powerful probe of the early Universe (e.g.~\cite{Caprini:2015zlo,Aggarwal:2020olq}). However, the limited extent in the range of frequencies associated with GWs from cosmological soliton formation often makes experimentally probing and constraining the parameter space of the theory challenging.

In this work we establish and explore a novel source of GWs associated with soliton isocurvature perturbations that are expected to be a
universal feature of cosmological scenarios involving soliton formation. Isocurvature perturbations correspond
to an unperturbed total energy density but inhomogeneities in the relative number densities of
different ``particle'' species. They are orthogonal to adiabatic (curvature) perturbations where the energy density perturbations do not affect number densities of particles.
Intriguingly, the resulting background of induced \textit{universal gravitational waves} (UGWs) sourced at second order by conversion of soliton isocurvature perturbations to curvature perturbations extends to much lower frequencies than GWs associated with soliton formation extensively discussed in the literature (e.g.~\cite{Caprini:2015zlo,Aggarwal:2020olq}).
This opens new routes for probing and constraining broad classes of cosmological theories across multiple GW frequency channels and enables us to test significantly more encompassing regions in their parameter space. As an explicit example, we discuss how this can lead to novel unique probes of cosmological production scenarios of axion-like particles (ALPs) where the ALP field fragments into oscillons.

{\it GWs from soliton isocurvature}.---
The mechanism for generating solitonic UGWs can be summarized as follows. In typical cosmological scenarios 
with soliton formation, the solitonic matter contributes sub-dominantly to the energy density during inflation and only weakly (e.g.~ gravitationally) interacts with the inflaton field. On length scales larger than the non-linear scale associated with the causal soliton formation process, the solitonic matter density field has a Poissonian power spectrum.
This Poissonian power spectrum ``tail'' plays the role of isocurvature perturbations appearing on sub-horizon scales at the end of soliton formation. As we demonstrate, conversion of these isocurvature perturbations into curvature perturbations sources at second order the induced UGW background. 
  
There are several key conceptual differences between the induced solitonic UGWs we discuss and the recently proposed production of an induced GW background from cold DM isocurvature~\cite{Domenech:2021and}.
First, UGWs are expected to be a universal feature of cosmological scenarios involving soliton formation and do not require additional assumptions such as existence of substantial initial isocurvature fluctuations in cold DM. Another important difference is that we are interested in isocurvature perturbations associated with sub-horizon scales when the Universe is at a finite age, which is distinct from Ref.~\cite{Domenech:2021and} that analyzed super-horizon isocurvature present from the beginning.  

We emphasize that an isocurvature perturbation can be generated not only on superhorizon scales but also on subhorizon scales. This is because the actual causal horizon size can be much smaller than the Hubble horizon. In the case considered in this paper, where the effective fluid composed of the solitons is pressureless, the causal horizon size is effectively zero when averaged over distances larger than the mean distance between the solitons. Nevertheless, this subhorizon isocurvature perturbation will eventually grow to be adiabatic as the effective pressureless fluid starts to dominate the universe. In this sense, namely from the point of view of the dynamics after generation, there is no difference between our case and the case of superhorizon isocurvature.

To discuss cosmological perturbations from primordial solitons in generality, we consider a perturbed flat Friedmann-Lemaıtre-Robertson-Walker (FLRW) Universe metric \cite{Kodama:1984ziu}
\Beq
ds^2=a^2(\tau)\left[-(1+2\Psi)d\tau^2 +(\delta_{ij}+2\Phi \delta_{ij}+h_{ij})dx^idx^j\right]\,,
\Eeq
where $a$ is the cosmic scale factor, $\tau$ is the conformal time, $\Psi$ is the lapse perturbation function, $\Phi$ is the curvature perturbation, and $h_{ij}$ is the tensor perturbation. We take the Universe to be dominated by a radiation fluid with density $\rho_r$ and also assume an energetically sub-dominant species of matter of density $\rho_\phi$ that results in formation of massive solitons at time $\tau_i$ through some mechanism, such as those mentioned earlier.

For sufficiently long-lived solitons initially comprising a minor contribution to the Universe's energy budget, they can lead to a stage of early matter domination that lasts until their decay into relativistic particles. The expansion history of this radiation-matter Universe during the soliton lifetime is given by $a/a_{\rm eq}=2\tau/\tau_\star+(\tau/\tau_\star)^2$, where 
$\tau_\star=2\sqrt{2}/\mathcal{H}_{\rm eq}$ and `eq' denotes the time of radiation-matter equality. 
Here $a H = \mathcal{H} = a'/a = (2/\tau)(1+\tau/\tau_\star)/(2+\tau/\tau_\star)$ and $\mathcal{H}_{\rm eq} = k_{\rm eq} = 2\sqrt{2}/\tau_{\star}$, with 
$k_{\rm eq}$ being the mode that enters the horizon at matter-radiation equality.
During this period, isocurvature perturbations are parameterized by \cite{Kodama:1986ud}
\Beq
S=\frac{\delta\rho_\phi}{\rho_\phi}-\frac{3}{4}\frac{\delta\rho_r}{\rho_r}\,,
\Eeq
and can source curvature perturbations. 

Following \cite{Domenech:2021and}, deep in the radiation domination epoch, i.e. when $\tau_i\leq\tau\ll\tau_\star$, the evolution equations capturing the linear generation of curvature perturbations by isocurvature perturbations, neglecting anisotropic stresses (with $\Phi\simeq-\Psi$), are given by
\Beq\label{eq:rdphi}
&\frac{d^2\Phi}{dx^2}+\frac{4}{x}\frac{d\Phi}{dx}+\frac{\Phi}{3}+\frac{1}{4\sqrt{2}\kappa x}\left[x\frac{d\Phi}{dx}+(1-x^2)\Phi-2S\right]=0\,,\\
&\frac{d^2S}{dx^2}+\frac{1}{x}\frac{dS}{dx}-\frac{x^2\Phi}{6}-\frac{1}{2\sqrt{2}\kappa}\left[\frac{dS}{dx}-\frac{x}{2}S-\frac{x^3}{12}\Phi\right]= 0\,,
\Eeq
where $k$ is the co-moving wavenumber, $x=k\tau$, $\kappa=k/\mathcal{H}_{\rm eq}\equiv k/k_{\rm eq}$. Here, the time coordinate $x$ separates superhorizon $(x \ll 1)$ and subhorizon $(x \gg 1)$ scales of the perturbation.
The approximate solution, considering
subhorizon scales with $x \gg 1$ and radiation domination regime with $x/\kappa = k_{\rm eq}\tau \ll 1$, is
\Beq\label{eq:sphisol}
&\Phi\simeq \frac{3S_{\bf k}(\tau_i)}{2\sqrt{2}\kappa} \frac{1}{x^3}\left[6+x^2-2\sqrt{3}x\sin\left(\frac{x}{\sqrt{3}}\right)-6\cos\left(\frac{x}{\sqrt{3}}\right)\right]\,,\\
&S\simeq S_{\bf k}(\tau_i)+\frac{3S_{\bf k}(\tau_i)}{2\sqrt{2}\kappa}\left[x+\sqrt{3}\sin\left(\frac{x}{\sqrt{3}}\right)-2\sqrt{3}{\rm Si}\left(\frac{x}{\sqrt{3}}\right)\right]\,,
\Eeq
where Si$(x)$ is the sine-integral function and the subscript ${\bf k}$ denotes the ${\bf k}$-dependence of the initial condition. We have verified that the approximate solutions in Eq.~\eqref{eq:sphisol}, derived for $\tau_i=0$ in Ref. \cite{Domenech:2021and}, are also attractor solutions for $x_i=k\tau_i>1$ and thus are applicable to our case of sub-horizon generation of isocurvature.

The curvature perturbation $\Phi$ generated from isocurvature can induce GWs at second order,
as we overview in Appendix.
Following expressions for induced GWs from curvature perturbations
~\cite{Kohri:2018awv,Domenech:2021ztg},
for Gaussian isocurvature perturbations the resulting induced GWs at time $x_c$ are given by \cite{Domenech:2021and}
\Beq\label{eq:GaussIso}
\Omega_{\rm GW,c}(k)=\frac{2}{3}\int_0^\infty &dv\int_{|1-v|}^{1+v}du\left[\frac{4v^2-(1-u^2+v^2)^2}{4uv}\right]^2\\
&\times\overline{I^2(x_c,k,u,v)}{\mathcal{P}_{S}(ku)}{\mathcal{P}_{S}(kv)}\,,
\Eeq
where we define the initial dimensionless power spectrum of isocurvature fluctuations, $\langle S_{\bf k}(\tau_i)S_{\bf k'}(\tau_i)\rangle=(2\pi^2/k^3)\mathcal{P}_S(k)\delta^{(3)}({\bf k}+{\bf k'})$. Here, $\overline{I^2}$ is the oscillation average of the square of the kernel. 
Unlike \cite{Domenech:2021and}, we consider the kernel for a finite initial time $x_i = k \tau_i$, since we are interested in sub-horizon modes at the time of soliton formation. See Appendix for computational details. Amplitude of UGWs will also depend on how prolonged is the soliton matter-dominated era.

The resulting GW spectrum today is then given by
\Beq
\Omega_{\rm GW,0}(k)= \Omega_{r,0}\left(\frac{g_{*}(T_{c})}{g_{*,0}}\right)\left(\frac{g_{* s}(T_{c})}{g_{* s,0}}\right)^{-4/3}\Omega_{\rm GW,c}(k)\,,
\Eeq
where $\Omega_{r,0}\simeq4.18\times 10^{-5}h^{-2}$ is the fractional density of radiation today~\cite{Planck:2018vyg}, 
and $g_{*}(T_{c})$ and $g_{* s}(T_{c})$ are the effective number of degrees of freedom in the energy and entropy densities at $T_c$, respectively. The present day values of the effective degrees of freedom are $g_{*,0}= 3.36$, $g_{* s,0} = 3.91$ and we assume the factors depending on them to be unity.

{\it ALP oscillons}.--- 
We illustrate the utility of our UGWs associated with soliton formation by calculating the induced GW production due to oscillons formed by an initially misaligned ALP field \cite{Hui:2016ltb,Cyncynates:2022wlq}. During inflation, the spectator ALP field is light and forms a condensate, which subsequently fragments into oscillons as the field becomes massive in the post-inflationary radiation-dominated Universe. As we demonstrate, UGWs from isocurvature contributions extend the frequency range of the GW signal by more than two orders of magnitude toward low frequencies, compared to GWs from oscillon formation, regardless of the ALP mass and decay constant. The strength of the signal depends on the fraction of the energy budget stored in the oscillons.
We note that UGWs are sourced by solitons formed in causal manner, and are expected to appear regardless of soliton originating nature (e.g. oscillons formed through mechanism of Ref.~\cite{Fukunaga:2019unq} or that of Ref.~\cite{Amin:2011hj}).

We consider the formation of oscillons from an ALP field $\phi$, governed by the canonical action and potential,
\Beq
S_\phi&=\int d^4x\sqrt{-g}\left[\frac{1}{2}(\partial\phi)^2-V(\phi)\right]\,;\\
&V(\phi)=m^2F^2\left[1-\cos\left(\frac{\phi}{F}\right)\right]\,.
\Eeq
During inflation $\phi$ is light with $m\ll H_{inf}$, and misaligned with $\phi_{inf}\sim F$. The ALP field starts oscillating after inflation when its mass becomes greater than the Hubble rate. This happens when the Universe is radiation dominated,  with $\rho_r\gg\rho_\phi$. 

\begin{figure}[t]
    \centering
\includegraphics[width=3.5in]{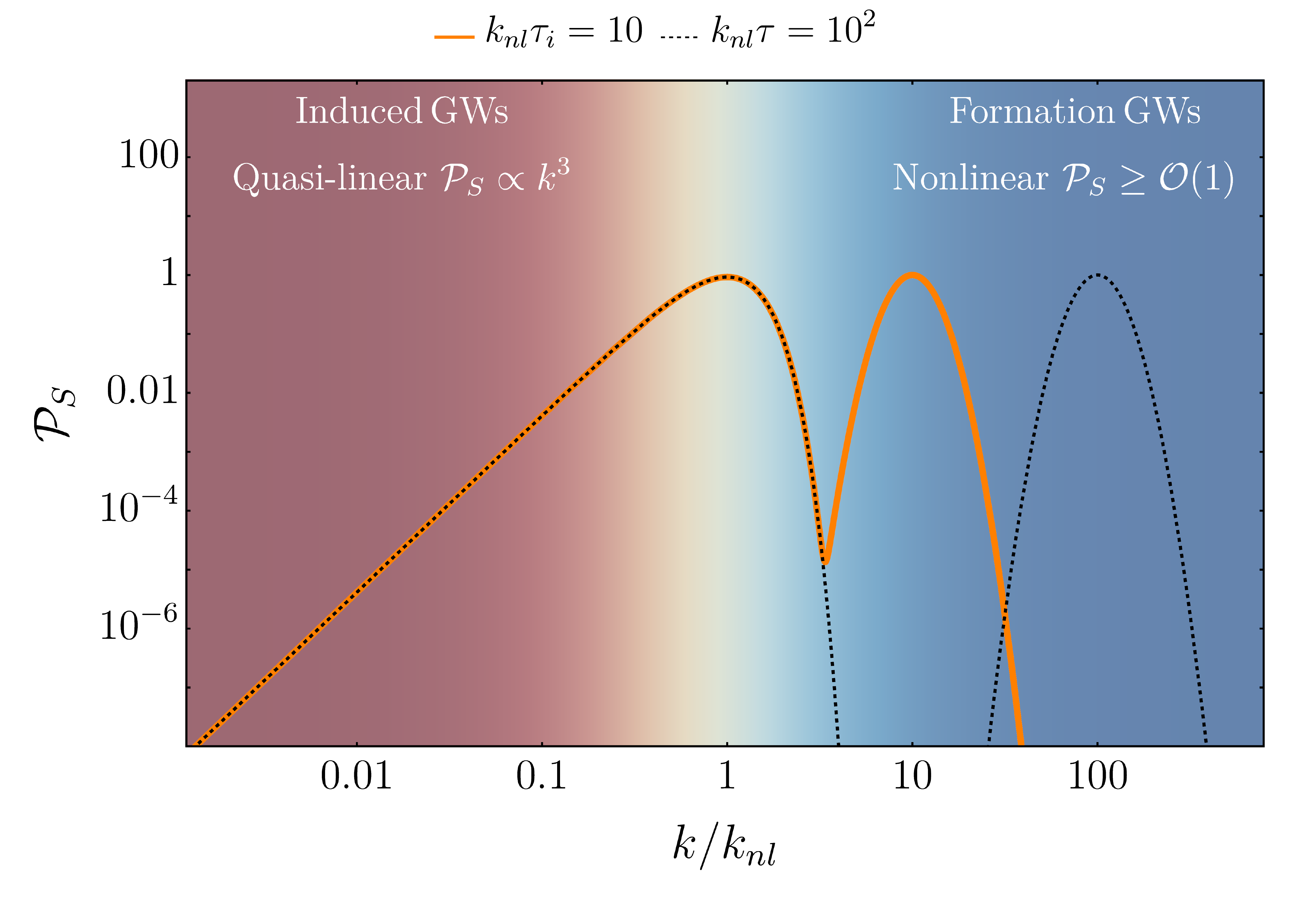}
\caption{Depiction of the oscillon isocurvature power spectrum at the time of formation (solid) and at the time of decay (dashed) for the fiducial parameters $k_{nl}\tau_i=10$, $k_{osc}\tau_i=10^2$, $k_{nl}=10^2k_{\rm eq}$. The UGWs is sourced by the Poissonian part of the spectrum $\propto k^3$, $k<k_{nl}$. The second peak on small scales \cite{Lozanov:2017hjm,Lozanov:2019ylm}, evolving towards higher $k$ is due to the individual oscillons. It leads to the generation of GWs only at the time of oscillon formation, but not afterwards.}
    \label{fig:PS}
\end{figure}

The oscillations of the $\phi$ condensate drive resonant instabilities of its initially small spatial inhomogeneities, resulting in rapid fragmentation into solitonic oscillon ``lumps''. The co-moving Hubble scale at the time of oscillon formation is $\mathcal{H}_i$.
Since $\delta\rho_\phi/\rho_\phi\gg\delta\rho_r/\rho_r$, the $\phi$ energy density power spectrum is approximately that of isocurvature $\simeq \mathcal{P}_S$ and has two peaks, see Fig.~\ref{fig:PS} for a qualitative illustration.

The first peak of the $\phi$ energy density power spectrum is located at a fixed physical wavenumber (thus, increasing co-moving wavenumber) corresponding to the fixed physical size of the oscillons $R_{osc}\sim m^{-1}$, located at
$k_{osc}(\tau)\sim a(\tau)m$.
The second peak is at a fixed co-moving wavenumber, corresponding to the fixed co-moving separation between the oscillons, $r_{nl}\sim 10a_im^{-1}$, located at
$k_{nl}\sim10^{-1}m/a_i={\rm const}\,$. Such oscillon separation scales are typically expected of generic potentials resulting in fragmentation of $\phi$ condensate~\cite{Lozanov:2017hjm}. The ``dip'' in power between the two peaks is due to transfer of matter, during oscillon formation, from these intermediate scales into the solitonic objects.
For $k<k_{nl}$, the $\phi$ energy density power spectrum is Poissonian ($\propto k^3$), since the formation of oscillon objects can be assumed as statistically independent events. Here we ignore the subsequent gravitational interactions of the oscillons, which is a good approximation in a radiation-dominated Universe.

We note that on much longer, super-horizon length scales, $k\ll k_{nl}$, probed by CMB, $\rho_\phi$ has a nearly scale-invariant power spectrum from inflation. This leads to constraints on the model parameters due to the CMB bounds on isocurvature. Specifically, the amplitude of the isocurvature perturbation of the spectator $\phi$ from quasi-de-Sitter inflation is given by~\cite{Agrawal:2018vin}
\Beq
\mathcal{P}_{\mathcal{S}}=\frac{H_{inf}^2}{\pi^2F^2\theta^2}\,,
\label{eq:Ps}
\Eeq
where $\theta=\phi_{inf}/F$. The isocurvature amplitude is constrained by the CMB observations~\cite{Planck:2018vyg} to be
$ \mathcal{P}_{\mathcal{S}}/A_s=\beta_{iso}<10^{-2}$,
where $A_s\simeq2.2\times10^{-9}$. Thus the axion decay constant $F$ is bounded as
\Beq
F\sim \frac{H_{inf}}{\sqrt{\pi A_s\beta_{iso}}}\gtrsim 10^5H_{inf}\,.
\Eeq
In what follows, we assume this bound is saturated, {\it i.e.}, $F\sim10^5H_{inf}$.

Since we assume the ALP field $\phi$ to be a spectator field during inflation, it follows that $H_{inf}\gg m$. It starts oscillating when $H\sim m$, and the oscillon formation occurs at $H_i\lesssim 10^{-1}m$ \cite{Fukunaga:2019unq,Amin:2011hj}. 
The initial total energy density $\rho_i$ is given by $\rho_i=3m_{pl}^2H_i^2$, which is equal to the radiation energy density $\rho_{r,i}$ with high accuracy. 
The initial oscillon density $\rho_{\phi,i}$ is related to the initial total energy density as $\rho_i/\rho_{\phi,i}\approx a_{eq}/a_i$.
Using $\rho_{\phi,i}\sim m^2F^2$ and $a_{eq}/a_i\approx \tau_{eq}/\tau_i=1/(k_{eq}\tau_i)$, the mass $m$ is related to the equality wavenumber $k_{\rm eq}$ as
$\rho_i/m^2F^2 \sim 1/k_{\rm eq}\tau_i$. This implies
\Beq \label{eq:malp}
m^2\sim m_{pl}^2\frac{k_{\rm eq}}{\mathcal{H}_i}\frac{H_i^2}{F^2}\,.
\Eeq

As fiducial values for the model parameters we choose $k_{\rm eq}\tau_i=10^{-1}$, $k_{nl}\tau_i=10$, $\beta_{iso}=10^{-2}$, which yields
$F\sim 10^{-3} m_{pl}$ and $H_{inf}\sim 10^{-8} m_{pl}$. From Eq.~\eqref{eq:malp} the axion mass scale $m$ is set by the choice of input parameters. For $H_i\sim10^{-19}m_{pl}$ (corresponding to the temperature of $\sim 10^9{\rm GeV}$), we have $m\sim10^{-18}m_{pl}$.

Hence, in our scenario the evolution of the Universe features several distinct stages as follows. The transition between the inflationary stage and reheating takes place at an energy scale $\rho_{inf}^{1/4}=(3m_{pl}H_{inf})^{1/2}$. Later on, the period when the subdominant $\phi$ starts oscillating and forms oscillons occurs when $\rho_i^{1/4}\sim (mm_{pl})^{1/2}$. The oscillons become dominant when $\rho_d^{1/4}\sim (k_{eq}\tau_i)^{1/2}\rho_i^{1/4}$. This is also the moment that we consider oscillons to decay.
For our fiducial parameters, $\rho_i^{1/4}\sim 10^9$~GeV and $\rho_d^{1/4}\sim 10^8$~GeV.

{\it Universal GWs from ALP oscillons}.--- After oscillon formation, i.e., for $\tau>\tau_i$, only the $k<k_{nl}$ part of the $\phi$ isocurvature contributes to the sourcing of GWs. For $k>k_{nl}$ the isocurvature power spectrum receives contributions only associated with individual oscillons (i.e. spherical soliton-like objects), with dimensionless energy-momentum tensor,
\Beq\label{eq:deltaTT}
\delta^{TT} = \frac{T^{TT}_\phi}{\rho_\phi+\rho_r}\sim\frac{T^{TT}_\phi}{\rho_\phi}\times\frac{\rho_\phi}{\rho_r}~,
\Eeq
where $T^{TT}_\phi$ is the $\phi$ transverse-traceless energy momentum tensor,
being $\delta^{TT}(k>k_{nl})\simeq0$.

\begin{figure}[t]
    \centering
\includegraphics[width=3.8in]{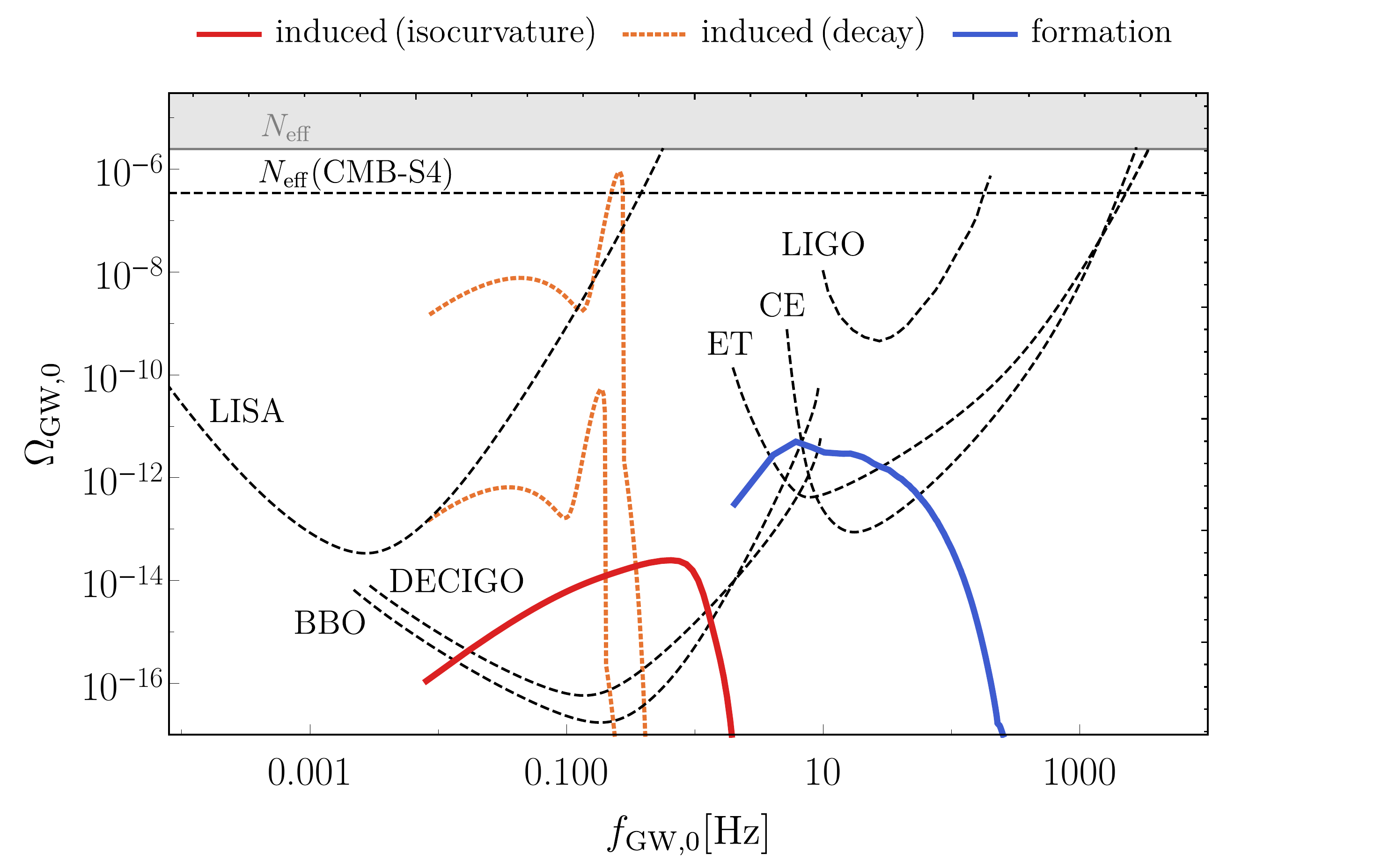}     \caption{The GW background from oscillon formation \cite{Lozanov:2019ylm} (blue), and the novel UGWs due to oscillon isocurvature (solid red) for the model parameters from Fig. \ref{fig:PS} and $m=10^{-18}m_{pl}$. The combination of the two GW signals presents the unique possibility of probing soliton physics across multiple wavebands with different GW experiments. The rapid transition from early oscillon matter-domination to radiation-domination epochs, accompanying the oscillon decay, will generically induce additional GWs at second order~\cite{Lozanov:2022yoy}. The lower and upper dashed orange curves, respectively, show the estimated induced GW signal from the linear part of $\mathcal{P}_S=(k/k_{nl})^3$, with $0<k<k_l$, for two different UV cut-offs, $k_l/k_{nl}=10^{-1}$ and $0.215$.}
    \label{fig:GWPS}
\end{figure}

If we assume that soon after the $\phi$ and the thermal bath energy densities become equal, around $\tau=1/k_{\rm eq}$, the oscillons evaporate into radiation, the strength of the stochastic GW background today is given in Fig. \ref{fig:GWPS}, and the frequency scale determined by $m$ is a free parameter. Here, the UGWs are computed considering constant infrared tail from isocurvature power spectrum as depicted in Fig.~\ref{fig:PS}. 

If we assume standard $\Lambda$CDM cosmological history after oscillon evaporation, the frequency today is given by
\Beq
f_0(k)\sim \frac{k/a_i}{\sqrt{H_i m_{pl}}} 10^{10}{\rm Hz}\,.
\Eeq
For $m\sim10^{-18} m_{pl}$, this gives
$f_0(k_{nl})\sim 1~{\rm Hz}$. For general ALP parameters, we observe that the frequency and the magnitude of the UGW signal scales as $f_0\propto m$ and $\Omega_{\rm GW}\propto (k_{eq}/k_{nl})^4$, respectively.

Furthermore, the transition of the Universe from the early oscillon matter domination stage to radiation domination will also generically induce GWs~\cite{Lozanov:2022yoy}. In a matter-dominated Universe, the scalar metric perturbations do not evolve with time. The sudden onset of their evolution at the transition to radiation domination can source GWs at second order \cite{Alabidi:2013lya,Inomata:2019ivs}. In Fig. \ref{fig:GWPS} we give this additionally induced GWs (assuming instantaneous decay of the oscillons into relativistic particles) sourced at second order, where a UV cut-off is assumed at the scale $k_l$ where the density perturbation at the matter-dominated stage becomes nonlinear, i.e., $\mathcal{P}_S(k)\lesssim (a_{eq}/a_{rad})^2$
for $k\leq k_l$, where $a_{rad}$ is the scale factor at the time of the matter-radiation transition. The thick orange curve is for $\mathcal{P}_S(k_l=10^{-1}k_{nl})=10^{-3}$,
and the dashed orange curve for $\mathcal{P}_S(k_l\approx0.215k_{nl})\approx10^{-2}$.
Note the sensitivity of the GW signal to the UV cut-off $k_l$ of the linear range of density perturbations. Further, unlike UGWs, the strength of the induced GWs from oscillon decays could be impacted by the details of matter-radiation transition~(e.g.~\cite{Inomata:2019ivs}). We note that induced GWs from matter-radiation era transition can also readily appear in other contexts, such as evaporating primordial black holes (e.g.~\cite{Domenech:2020ssp,Domenech:2021wkk,Papanikolaou:2020qtd}). We stress that the magnitude of the (orange dashed) curve depicting induced GWs associated with oscillon decays in Fig.~\ref{fig:GWPS} has model dependency and can be significantly affected if oscillons do not suddenly decay. Oscillon decays taking place over a Hubble time scale will result in a suppressed signal, however, the Universal GWs (the red curve) that are unrelated to oscillon decays will persist regardless and hence would constitute the dominant contribution at these frequencies.

For comparison we also give the GW power spectrum generated during oscillon formation \cite{Lozanov:2019ylm}. Its peak frequency is close to $k_{osc}$ and power,
\Beq
\Omega_{\rm GW}(f\sim k_{osc}(\tau_i))\sim (\delta^{TT})^2\frac{\mathcal{H}_i^2}{k_{osc}^2(\tau_i)}\,,
\Eeq
where $\delta^{TT} \lesssim \mathcal{O}(1)\times (k_{\rm eq}\tau_i)$.

We note that for an additional assumption of prolonged matter-domination era could result in additional features on the low frequency tail of the GW spectrum~\cite{Hook:2020phx}.
The isocurvature GWs are expected to scale as $k^3$ for $k$ approaching 0~\cite{Cai:2019cdl}. 
This additional isocurvature $k^3$ tail should lie above the $k^3$ signal discussed in Ref.~\cite{Hook:2020phx}, where the isocruvature source was neglected.

{\it Universal GWs from Solitons}.---The induced GWs in the oscillon scenario presented above are a {\it generic} prediction of all models of the primordial Universe, featuring soliton formation. The presence of the induced GW signal is independent of the nature of the solitons.

This is because, in all soliton scenarios, the density of the matter field comprising the solitons (and sourcing the GWs), has a universal shape. On the shortest scales, $k_s$, associated with the individual objects ($k_{osc}$ in the case of oscillons), the matter field is overdense, $\mathcal{P}_S(k_s)\gg \mathcal{O}(1)$. On longer scales, bounded by the typical distance scale between neighbouring objects, $k_{nl}$, the density field is nonlinear $\mathcal{P}_S(k\gtrsim k_{nl})\sim \mathcal{O}(1)$. On much longer scales, the spatial inhomogeneities in the matter can be treated as linear perturbations, $\mathcal{P}_S(k\ll k_{nl})\ll \mathcal{O}(1)$. In the intermediate range of $k$, where the density field transitions from linear to nonlinear, the power spectrum should take a Poissonian form due to the causal nature of the soliton formation mechanism, $\mathcal{P}_S(k\lesssim k_{nl})\sim (k/k_{nl})^3$.

The main focus in the literature has been on the generation of GWs by the nonlinear part of the source, $\mathcal{P}_S(k\gtrsim k_{nl})$. The numerical lattice studies of the formation and evolution of e.g., domains walls, cosmic string networks, monopoles, textures, Q-balls, oscillons etc., capture in great detail the dynamics on the nonlinear scales, which indeed provides the leading contribution to the GW signal in terms of energy. However, due to limitations in their dynamic range, lattice simulations often do not cover the quasi-linear large-scale Poissonian part of the source. 

In this work we put forth, and explicitly demonstrate realization in the case of ALP oscillons, that the Poisson tail of the spatial distribution can significantly extend the GW power spectrum to much lower frequencies.
This property is shared by all localized solitons, e.g., monopoles, Q-balls, oscillons, etc. 

It is known that the brief nonlinear dynamics of formation gives rise to a single-peak GW power spectrum centered on a nonlinear scale $k\sim k_s\gg k_{nl}$. Subsequently, according to our analysis, the Poissonian part of the density field at quasi-linear scales $k\lesssim k_{nl}$ continuously sources GWs, significantly extending the infrared tail of the GW spectrum.

Configurations of non-localized solitons, e.g., cosmic string networks, continuously source GWs during their formation and evolution. The nonlinearly sourced GW spectrum can be extremely broad (even scale-invariant). We expect that the GW signal on quasi-linear scales receives a contribution from the large scale distribution of solitonic configurations in analogy to the case of localized solitons.

{\it Conclusions}.--- Cosmological solitonic objects such as oscillons, Q-balls, monopoles appear in many distinct theories of the early Universe. We found that such scenarios are generically accompanied by production of novel induced \textit{universal gravitational waves} (UGWs), associated with soliton isocurvature perturbations. Intriguingly, universal gravitational waves appear at much lower frequencies than the gravitational waves associated with characteristic scales of soliton formation. In the case of ALP oscillons, UGWs extend the gravitational wave spectrum by orders of magnitude to lower frequencies and can be observable in upcoming experiments. 
UGWs establish novel probes and signatures for many distinct classes of cosmological theories leading to soliton formation, substantially improving and broadening the power of GW experiments and opportunities to explore the early Universe.

{\it Acknowledgments}.---We thank Mustafa Amin, Guillem Domenech for helpful discussions.
This work is supported by World Premier International
Research Center Initiative (WPI), MEXT, Japan.
This work is also supported in part by the JSPS KAKENHI grant Nos. JP19H01895 (M.S.), JP20H04727 (M.S.), JP20H05853 (M.S.), JP23K13109 (V.T.) and JP24K00624 (M.S.).

\appendix

\section{Soliton Induced Universal Gravitational Waves}
\label{sec:indgws}

At second order in perturbations, scalar modes source tensor modes (for review see e.g.~\cite{Domenech:2021ztg}). The equation of motion for tensor modes is
\Beq\label{eq:GWs}
h_{ij}''+2\mathcal{H}h'_{ij}+\Delta h_{ij}=\mathcal{P}_{ij}{}^{ab}\mathcal{S}_{ab}\,,
\Eeq
where $\mathcal{P}_{ij}$ is the transverse-traceless projector (see e.g.~\cite{Domenech:2021ztg,Papanikolaou:2020qtd}) and the quadratic scalar source is
\Beq
\label{eq:quadsource}
\mathcal{S}_{ij}\simeq4\partial_i\Phi\partial_j\Phi+2\partial_i\left[\frac{\Phi'}{\mathcal{H}}
+\Phi\right]\partial_j\left[\frac{\Phi'}{\mathcal{H}}+\Phi\right]\,.
\Eeq
We note that in Eq.~\eqref{eq:quadsource} we suppressed the contributions from relative velocity perturbations, which are suppressed on sub-horizon scales that we are interested in.

We discuss the details and kernel of induced UGWs calculations for solitons, following Ref.~\cite{Domenech:2021and} and highlighting important differences with their calculations. Repeating Eq.~(5) of the main text, for Gaussian isocurvature perturbations the resulting induced GWs at time $x_c$ are given by 
\begin{align}\label{eq:GaussIso}
\Omega_{\rm GW,c}(k)=&~\frac{2}{3}\int_0^\infty dv\int_{|1-v|}^{1+v}du\left[\frac{4v^2-(1-u^2+v^2)^2}{4uv}\right]^2 \notag\\~&\times\overline{I^2(x_c,k,u,v)}{\mathcal{P}_{S}(ku)}{\mathcal{P}_{S}(kv)}\,,
\end{align}
where $\overline{I^2}$ is
the oscillation average of kernel
\Beq\label{eq:Kernel}
I(x_c,k,u,v)\equiv x_{c}\int_{x_i(k)}^{x_c}d\tilde{x} G(x_{c},\tilde{x})f(\tilde{x},k,u,v)\,,
\Eeq
involving the Green function of GWs, see eq. \eqref{eq:GWs}.
In a radiation dominated Universe
\Beq
G(x,\tilde{x})= \frac{a(\tilde{x})}{a(x)}(\sin(x)\cos(\tilde{x})-\cos(x)\sin(\tilde{x}))\,,
\Eeq
and the scalar source term
\begin{widetext}
\begin{align}\label{eq:fapp}
\kappa^2f(x,u,v)\simeq&\frac{9}{8 u^2 v^2 x^2}+\frac{27 \left(u^2+v^2\right)}{4 u^4 v^4 x^4}+\frac{243}{2 u^4 v^4 x^6}+\frac{27}{2}\left(\frac{\left(2 u^2-v^2\right)}{2u^4 v^4
   x^4}-\frac{9}{u^4 v^4 x^6}\right) \cos \left(u\frac{x}{\sqrt{3}}\right)\notag\\
   &-\frac{27}{2}\left(\frac{\left(u^2-2 v^2\right)}{2 u^4 v^4
   x^4}+\frac{9}{u^4 v^4 x^6}\right) \cos \left({v
   \frac{x}{\sqrt{3}}}\right)-\frac{9 \sqrt{3}}{2}\left(\frac{1}{2 u^2
   v^3 x^3}+\frac{9}{u^4v^3x^5}\right) \sin \left({v\frac{x}{\sqrt{3}}}\right)\notag\\
   &-\frac{9 \sqrt{3}}{2}\left(\frac{1}{2 u^3 v^2 x^3}+\frac{9
   }{u^3 v^4 x^5}\right) \sin \left(u\frac{x}{\sqrt{3}}\right)+\frac{9 \sqrt{3}(u-v)}{4}\left(\frac{1}{u^3 v^3 x^3}+\frac{9}{u^4v^4x^5}\right) \sin \left({
   (u-v)}\frac{x}{\sqrt{3}}\right)\notag\\
   &-\frac{9 \sqrt{3}(u+v)}{4}\left(\frac{1}{ u^3 v^3 x^3}-\frac{9}{u^4 v^4 x^5}\right) \sin \left({
   (u+v)}\frac{x}{\sqrt{3}}\right)+\frac{9}{4}\left(\frac{1}{2 u^2 v^2 x^2}-\frac{3 \left(u^2-3 u v+v^2\right)}{u^4v^4x^4}+\frac{27}{u^4v^4x^6}\right)\notag\\
   &\times\cos\left({(u-v)}\frac{x}{\sqrt{3}}\right)+\frac{9}{4}\left(\frac{1}{2 u^2 v^2 x^2}-\frac{3 \left(u^2+3uv+v^2\right)}{ u^4 v^4
   x^4}+\frac{27}{u^4v^4x^6}\right) \times\cos\left({(u+v)}\frac{x}{\sqrt{3}}\right)\,.
\end{align}
\end{widetext}
The oscillation averaged kernel, for sufficiently large $x_c$, reduces to
\Beq\label{eq:SqKernel}
\overline{I^2(x_c,k,u,v)}\simeq\frac{1}{2}\left[I_{cos,\infty}^2(k,u,v)+I_{sin,\infty}^2(k,u,v)\right]\,,
\Eeq
where since we assume radiation domination we have
\Beq\label{eq:kernelcossin1}
I(x,k,u,v)=I_{cos}(x,k,u,v)\sin x-I_{sin}(x,k,u,v)\cos x\,,
\Eeq
and
\Beq\label{eq:kernelcossin2}
I_{cos/sin}(x,k,u,v)\equiv\int_{x_i(k)}^{x}d\tilde{x}\tilde{x}\left\{\begin{aligned}
	&\cos \tilde x\\
	&\sin \tilde x
	\end{aligned}
	\right\} f(\tilde{x},k,u,v)\,,
\Eeq
and $I_{cos/sin,\infty}(k, u, v)\equiv\lim_{x_c \rightarrow \infty}I_{cos/sin}(x_c,k,u,v)$. 

Unlike the superhorizon CDM isocurvature scenario \cite{Domenech:2021and} in which $x_i(k)$ was set to zero, in our case of late-time generation of sub-horizon soliton isocurvature the lower bound of the integral in Eq.~\eqref{eq:Kernel} is $k$-dependent and typically greater than unity, preventing the further simplification of the analytic expressions for the integral kernel.

\bibliography{IGWSIsoOscillons}

\end{document}